\documentclass[amsmath,amssymb,superscriptaddress,nobalancelastpage,prl,twocolumn,preprintnumbers,floatfix]{revtex4-1}

\usepackage[paperwidth=8.375in,paperheight=10.875in,left=0.55in,right=0.55in,top=0.55in,bottom=0.60in,columnsep=0.22in]{geometry}
\usepackage{bm}
\usepackage{braket}
\usepackage{diagbox}
\usepackage{graphicx}
\usepackage{hyperref}
\hypersetup{colorlinks,linkcolor=blue,urlcolor=blue,citecolor=blue}
\usepackage{siunitx}
\usepackage{ulem}
\usepackage{varioref}
\usepackage{xr-hyper}
\usepackage{xcolor}
\colorlet{ORANGE}{orange}
\definecolor{TIPred}{HTML}{E24B43}
\setcitestyle{super,sort&compress}

\newcommand{\TIPsection}[1]{\par\medskip\noindent{\color{TIPred}\bfseries\MakeUppercase{#1}}\par\smallskip}
\newcommand{\TIPsubsection}[1]{\par\smallskip\noindent{\color{TIPred}\bfseries #1}\par}

\providecommand{\Endparasplit}{}

\begin{document}

\setlength{\parindent}{1em}

\author{Xunyang~Hong}
\affiliation{Department of Physics, The Chinese University of Hong Kong, Hong Kong SAR 999077, China}
\affiliation{Physik-Institut, Universit\"{a}t Z\"{u}rich, Winterthurerstrasse 190, Z\"{u}rich CH-8057, Switzerland}

\author{Yujie~Yan}
\affiliation{Department of Physics, The Chinese University of Hong Kong, Hong Kong SAR 999077, China}

\author{Leonardo~Martinelli}
\affiliation{Physik-Institut, Universit\"{a}t Z\"{u}rich, Winterthurerstrasse 190, Z\"{u}rich CH-8057, Switzerland}

\author{Izabela~Bia\l{}o}
\affiliation{Physik-Institut, Universit\"{a}t Z\"{u}rich, Winterthurerstrasse 190, Z\"{u}rich CH-8057, Switzerland}

\author{Karin~von~Arx}
\affiliation{Physik-Institut, Universit\"{a}t Z\"{u}rich, Winterthurerstrasse 190, Z\"{u}rich CH-8057, Switzerland}

\author{Jaewon~Choi}
\affiliation{Diamond Light Source, Harwell Campus, Didcot, Oxfordshire OX11 0DE, United Kingdom}
\affiliation{Department of Physics, Korea Advanced Institute of Science and Technology, 291 Daehak-ro, Daejeon 34141, South Korea}
   
\author{Yasmine~Sassa}
\affiliation{Department of Applied Physics, KTH Royal Institute of Technology, Stockholm SE-106 91, Sweden}

\author{Sunseng~Pyon}
\affiliation{Department of Applied Physics, The University of Tokyo, Tokyo 113-8656, Japan}

\author{Tomohiro~Takayama}
\affiliation{Max Planck Institute for Solid State Research, Stuttgart 70569, Germany}

\author{Hidenori~Takagi}
\affiliation{Max Planck Institute for Solid State Research, Stuttgart 70569, Germany}
\affiliation{Department of Physics, The University of Tokyo, Tokyo 113-0033, Japan}

\author{Chafic~Fawaz}
\affiliation{Institut N\'eel, CNRS and Universit\'e Grenoble Alpes, Grenoble 38042, France}

\author{Laura~Chaix}
\affiliation{Institut N\'eel, CNRS and Universit\'e Grenoble Alpes, Grenoble 38042, France}

\author{Mehdi~Frachet}
\affiliation{Institut N\'eel, CNRS and Universit\'e Grenoble Alpes, Grenoble 38042, France}

\author{Igor~Vinograd}
\affiliation{Univ. Grenoble Alpes, CNRS, Univ. Toulouse, INSA-T, LNCMI-EMFL, Grenoble UPR3228, France}

\author{Flora~Yakhou}
\affiliation{European Synchrotron Radiation Facility, 71 Avenue des Martyrs, Grenoble F-38043, France}

\author{Kaya~Vogler}
\affiliation{Physik-Institut, Universit\"{a}t Z\"{u}rich, Winterthurerstrasse 190, Z\"{u}rich CH-8057, Switzerland}

\author{Dan~Zhao}
\affiliation{Hefei National Laboratory for Physical Sciences at the Microscale, University of Science and Technology of China, Hefei 230026, China}

\author{Yanbing~Zhou}
\affiliation{Hefei National Laboratory for Physical Sciences at the Microscale, University of Science and Technology of China, Hefei 230026, China}

\author{Tao~Wu}
\affiliation{Hefei National Laboratory for Physical Sciences at the Microscale, University of Science and Technology of China, Hefei 230026, China}
\affiliation{CAS Key Laboratory of Strongly-coupled Quantum Matter Physics, Department of Physics, University of Science and Technology of China, Hefei 230026, China}
\affiliation{Collaborative Innovation Center of Advanced Microstructures, Nanjing University, Nanjing 210093, China}

\author{Xianhui~Chen}
\affiliation{Hefei National Laboratory for Physical Sciences at the Microscale, University of Science and Technology of China, Hefei 230026, China}
\affiliation{CAS Key Laboratory of Strongly-coupled Quantum Matter Physics, Department of Physics, University of Science and Technology of China, Hefei 230026, China}
\affiliation{Collaborative Innovation Center of Advanced Microstructures, Nanjing University, Nanjing 210093, China}

\author{Marc-Henri~Julien}
\affiliation{Univ. Grenoble Alpes, CNRS, Univ. Toulouse, INSA-T, LNCMI-EMFL, Grenoble UPR3228, France}

\author{Mirian~Garcia-Fernandez}
\affiliation{Diamond Light Source, Harwell Campus, Didcot, Oxfordshire OX11 0DE, United Kingdom}

\author{Ke-Jin~Zhou}
\affiliation{Diamond Light Source, Harwell Campus, Didcot, Oxfordshire OX11 0DE, United Kingdom}
   
\author{Johan~Chang$^{*}$}
\affiliation{Physik-Institut, Universit\"{a}t Z\"{u}rich, Winterthurerstrasse 190, Z\"{u}rich CH-8057, Switzerland}

\author{Qisi~Wang$^{*}$}
\affiliation{Department of Physics, The Chinese University of Hong Kong, Hong Kong SAR 999077, China}
\affiliation{State Key Laboratory of Quantum Information Technologies and Materials, The Chinese University of Hong Kong, Hong Kong SAR 999077, China}

\title{Unveiling the nature of collective charge excitations in a cuprate
superconductor}

\maketitle

\noindent\textbf{*Correspondence:} johan.chang@physik.uzh.ch (J.C.); qwang@cuhk.edu.hk (Q.W.)

\TIPsection{ABSTRACT}

\noindent
Emergent symmetry breakings in condensed matter systems are often intimately linked to collective excitations. For example, the intertwined spin-charge stripe order in cuprate superconductors is associated with spin and charge excitations. While the collective behavior of spin excitations is well established, the nature of charge excitations remains to be understood. Here we present a high-resolution resonant inelastic x-ray scattering (RIXS) study of charge excitations in the stripe-ordered cuprate La$_{1.675}$Eu$_{0.2}$Sr$_{0.125}$CuO$_4$. The RIXS spectra consist of both charge and phonon excitations around the charge ordering wave vector. By modeling the momentum-dependent phonon intensity, the charge-excitation spectral weight is extracted over a wide energy range. As such, we reveal the highly dispersive nature of the charge excitations, with an energy scale comparable to the spin excitations. Since charge order and superconductivity in cuprates are possibly driven by the same electronic correlations, determining the interaction strength underlying charge order is essential to establishing a comprehensive microscopic model of high-temperature superconductivity.

\TIPsection{INTRODUCTION}

Correlated electron systems often exhibit a variety of collective excitations that encode the nature of their ground states. 
For example, in cuprate superconductors, observations of high-energy spin excitations
reveal strong magnetic correlations and large exchange couplings that persist across  the entire doping phase diagram.\cite{dean_persistence_2013} These findings  have led to the belief that spin fluctuations are primarily responsible for the formation of Cooper pairs in these materials.\cite{ScalapinoRMP2012,monthoux_superconductivity_2007}
Besides magnetic correlations, experiments have established the universal presence of charge correlations in hole-doped cuprates,\cite{ghiringhelli_long-range_2012, peng_direct_2016, da_silva_neto_ubiquitous_2014, chaix_bulk_2022, tabis_synchrotron_2017}  with a possible link to superconductivity and the pseudogap phase.\cite{caprara_dynamical_2017, perali_d-wave_1996, yue_pseudogap_2024, yu_unusual_2020, kivelson_electronic_1998}
While static charge order is found to compete with superconductivity,\cite{chang_tuning_2008} its bosonic excitations could potentially generate attractive interactions between electrons and promote superconductivity.\cite{yu_unusual_2020,wang_enhancement_2015}
Recent studies further suggest charge fluctuations to be responsible for the strange metal behavior,\cite{arpaia_signature_2023} and the pseudogap formation.\cite{caprara_dynamical_2017, yue_pseudogap_2024,yu_unusual_2020} These findings have rendered charge dynamics a crucial factor for understanding the low-energy physics of cuprates.
In particular, determining the interaction strength behind charge order is crucial for establishing the effective model that describes the complex phase diagram of cuprates.\cite{davis_concepts_2013, sachdev_bond_2013,zheng_stripe_2017,huang_numerical_2017}
However, it has been a challenging experimental task to directly resolve collective charge excitations. How the charge excitations influence the electronic properties of cuprates, therefore, remains an outstanding question.
\Endparasplit

With improvement in energy resolution, resonant inelastic x-ray scattering (RIXS) has been established as an effective probe of low-energy excitations.\cite{AmentRMP2011}
The enhanced sensitivity to charge correlations is especially useful for studies of strongly correlated systems.
Recent RIXS studies on several cuprate systems have revealed anomalies of phonon dispersions and intensities near the charge ordering wave vector,\cite{li_multiorbital_2020, lin_strongly_2020, huang_quantum_2021, chaix_dispersive_2017, li_prevailing_2023, peng_enhanced_2020} implying a possible charge excitation mode that intersects the phonon modes.
Notably, a low-energy excitation mode with a characteristic energy scale of $\sim$$10$~meV has been observed near the charge ordering wave vector.\cite{huang_quantum_2021,arpaia_signature_2023, martinelli_decoupling_2025, li_prevailing_2023} This low-energy mode, identified as charge density fluctuations, has been proposed to account for the strange metal behavior.\cite{arpaia_signature_2023}
Yet, due to the interference between the weak charge excitations and various phonon modes, it has been difficult to extract the charge excitation spectral weight. As such, essential information including the dispersion and energy scale of the charge excitation mode remains unknown.

\begin{sloppypar}
Here, we employ oxygen $K$-edge RIXS to investigate the low-energy excitations in La$_{1.8-x}$Eu$_{0.2}$Sr$_{x}$CuO$_4$ with $x=0.125$ (Eu-LSCO), which exhibits a strong charge order.\cite{wang_high-temperature_2020}
By modeling the RIXS phonon cross sections, we extract the spectral weight of charge excitations as a function of momentum and energy transfers $S(\mathbf{q},\omega)$. 
At low temperature ($T=21$~K) the charge excitation mode emerges from the ordering wave vector and disperses outwards with a velocity of $v_c \sim 800$ meV \AA ~up to at least $\sim$60~meV.
Such an energy scale is comparable to the paramagnon bandwidth, implying the connection to the spin order as proposed within the framework of fluctuating stripes. The highly dispersive character reveals the collective nature of charge excitations, which may have a profound impact on the superconducting and normal state properties.
\end{sloppypar}

\TIPsection{MATERIALS AND METHODS}

\begin{figure*}[!t]
\centering
\includegraphics[width=\linewidth]{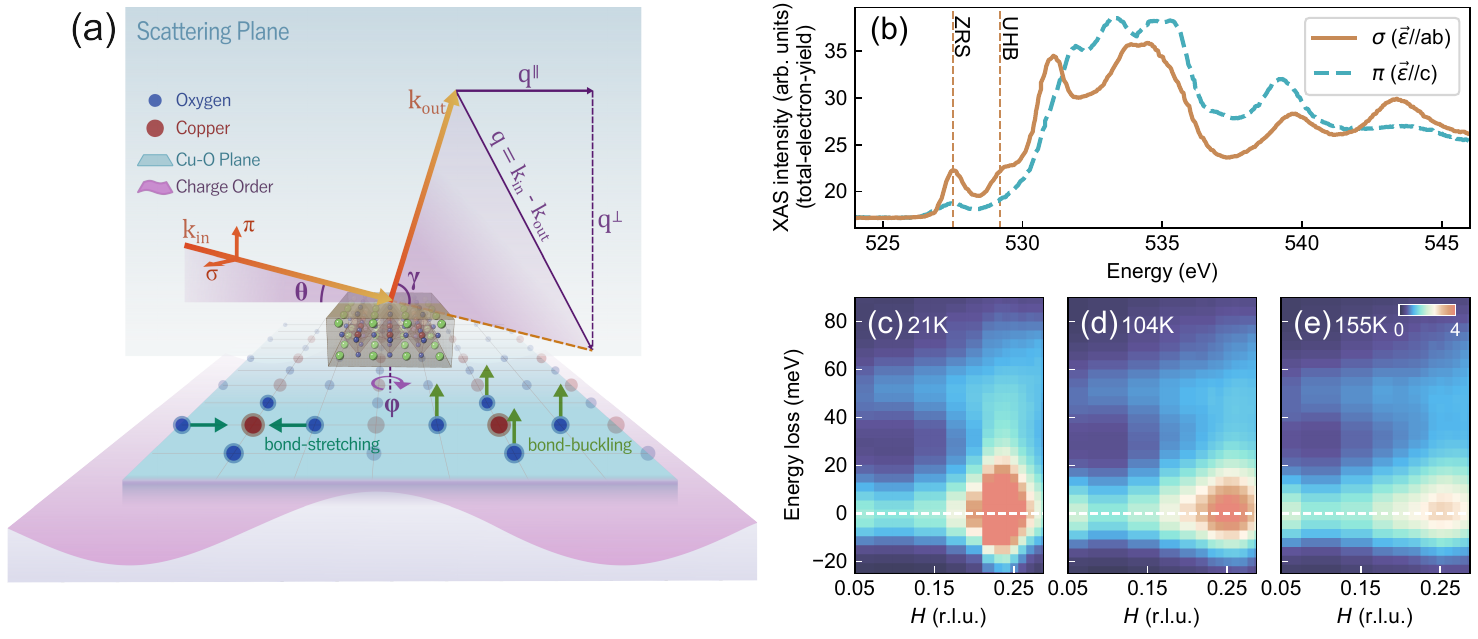}
\caption{
    Crystal structure, phonon modes, XAS, and RIXS intensity map (a) Schematic of the RIXS experiment. Momentum transfer $\mathbf{q}$ between the incident ($\mathbf{k}_{\rm{in}}$) and scattered ($\mathbf{k}_{\rm{out}}$) photons is projected into components within ($\mathbf{q}_{//}$) and perpendicular to ($\mathbf{q}_{\perp}$) the Cu-O planes. Green arrows illustrate the bond-stretching (in-plane) and bond-buckling (out-of-plane) phonon modes.
    (b) XAS spectra (measured by total-electron-yield) collected with a grazing-incidence angle of $\theta=20^{\circ}$ using $\sigma$ and $\pi$ light polarizations. (c--e) RIXS intensity map as a function of energy loss and momentum transfer at indicated temperatures.  White dashed lines represent the zero energy loss.}
\label{fig:fig1}
\end{figure*}

\TIPsubsection{RIXS experiments}
High-quality Eu-LSCO single crystals were grown using the floating zone method and studied at the I21 beamline of Diamond Light Source.\cite{zhou_i21_2022}
RIXS experiments were carried out at the O $K$-edge resonance ($\sim$$527.5$~eV). 
Grazing-incidence geometry with linear vertical ($\sigma$) incident light polarization was employed to enhance the cross sections of charge and phonon excitations. 
The energy resolution, determined from the elastic scattering of amorphous carbon, was measured to have a full-width-at-half-maximum (FWHM) of 22.5 meV.
The wave vector $\mathbf{q}$ at $(q_{x},q_{y},q_{z})$ was defined as $(H,K,L)=(q_{x}a/2\pi,q_{y}b/2\pi,q_{z}c/2\pi)$ in reciprocal lattice units (r.l.u.) using pseudo-tetragonal notation, with $a\approx b\approx3.79$~\AA~and $c\approx13.1$~\AA.
The scattering angle was fixed at $\gamma=154^{\circ}$. The magnitude and direction of the in-plane momentum transfer $\mathbf{q}_{\parallel}=(H,H \tan \phi)$ are controlled by the x-ray incident angle $\theta$ and sample azimuthal angle $\phi$, respectively (Figure~\ref{fig:fig1}a). 
To avoid the overwhelming elastic scattering at the charge ordering wave vector, the data were taken with $\phi = 4^{\circ}$.\cite{wang_charge_2021}
RIXS intensities were normalized to counting time.
Additionally, we repeated all analyses using data normalized to the weight of the $dd$ excitations between 1.0~eV and 3.3~eV, as in Refs.\cite{ghiringhelli_long-range_2012,lin_strongly_2020,wang_charge_2021,arpaia_signature_2023} Both normalization methods yield consistent conclusions (see Supplemental Information).
Additional copper $L$-edge RIXS data on La$_{1.8-x}$Eu$_{0.2}$Sr$_{x}$CuO$_4$ with $x=0.26$ (Eu-LSCO, $x=0.26$), in the Supplemental Information, were measured at the ID32 beamline of European Synchrotron Radiation Facility (ESRF), with $\gamma=149.5^{\circ}$ and resolution of 55 meV.

\TIPsubsection{Phonon simulation}
Inelastic neutron scattering (INS) and inelastic x-ray scattering (IXS) phonon cross sections are proportional to the dynamical structure factor.\cite{shirane_neutron_2002} In contrast, the RIXS phonon cross section depends on resonance effects, reflecting the momentum-dependent electron-phonon coupling.\cite{AmentRMP2011}
The RIXS phonon intensities are calculated by a diagrammatic method:\cite{devereaux_directly_2016} 
\begin{equation}
I_\nu(\mathbf{q}, \Omega) = \sum_{\mathbf{k},\mathbf{k}^\prime} G[g]\,P[\phi]\,E[\varepsilon_{\mathbf{k}}, D_\nu],
\end{equation}
where $I_\nu$ is the RIXS intensity; $\mathbf{q}$ and $\Omega$ are the momentum transfer and the energy loss, respectively; the subscript $\nu$ indicates the phonon mode; and $\mathbf{k}$ is the momentum variable on the electron energy band $\varepsilon_{\mathbf{k}}$. 
In the equation, $G[g]$ is a function of electron-phonon coupling $g(\mathbf{k}, \mathbf{q})$, $P[\phi]$ is a function of orbital projection function $\phi_{2p_x}(\mathbf{k})$, $E[\varepsilon_{\mathbf{k}}, D_\nu]$ is a function of energy band $\varepsilon_{\mathbf{k}}$ and phonon propagator $D_\nu(\mathbf{q})$ (see Supplemental Information for detailed descriptions).

\TIPsection{RESULTS}
X-ray absorption spectroscopy (XAS) measurements were performed across the oxygen $K$-edge resonance
using $\sigma$ and $\pi$ polarizations with grazing-incidence geometry (Figure~\ref{fig:fig1}b). The 
pre-edge at $\sim$$527.5$~eV is commonly interpreted as a Zhang-Rice singlet (ZRS) state,\cite{chen_electronic_1991, lee_generic_2022} where an oxygen $1s$ electron is excited into the unoccupied oxygen $2p$ ligand hole state ($3d^9\underline{L} \rightarrow \underline{1s}3d^9$). The peak at $\sim$$529$~eV originates from the excitation of a $1s$ core electron to the upper Hubbard band (UHB, $3d^9 \rightarrow \underline{1s}3d^{10}$). Both the ZRS and UHB peaks are enhanced with $\sigma$ light polarization, indicating a dominant in-plane electronic nature.

\begin{figure}[!t]
    \centering
    \includegraphics[width=0.92\linewidth]{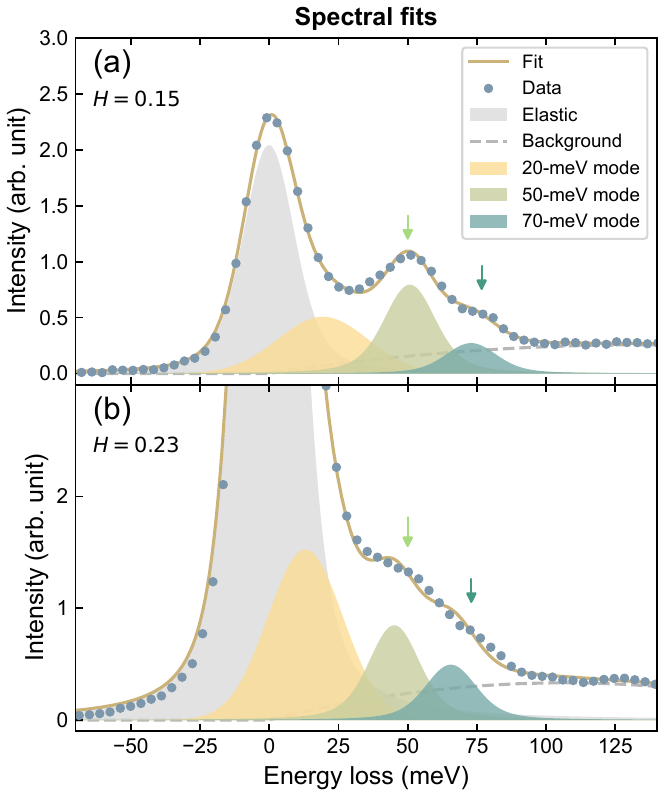}
    \caption{Spectral fits of the low-energy RIXS excitations (a, b) Representative RIXS spectra recorded away from and near the charge ordering wave vector at base temperature (21~K). The brown solid line is a fit including elastic scattering (grey), a charge mode (yellow), and two phonon modes (light green and dark green) on top of a background (grey dashed line). The arrows indicate the energies used for the simulations of the corresponding phonons (see main text).}
    \label{fig:fig2}
\end{figure}

RIXS measurements were performed at the ZRS peak energy ($\sim$$527.5$~eV) which has enhanced sensitivity to the doped holes. RIXS intensity maps as a function of in-plane momentum transfer $\mathbf{q}=(H,0)$ and energy loss at three selected temperatures are shown in Figures~\ref{fig:fig1}c--e. 
Consistent with previous resonant and non-resonant x-ray scattering studies,\cite{wang_high-temperature_2020, wang_uniaxial_2022, tabis_synchrotron_2017, miao_high-temperature_2017, miao_formation_2019, lin_strongly_2020,huang_quantum_2021, choi_unveiling_2022, christensen_bulk_2014} the charge order reflection at $\mathbf{q_\text{CO}}=(q_\text{CO},0)$, with $q_\text{CO}\approx0.23$, weakens with increasing temperature.
The broad feature in momentum at $155$~K indicates a dynamical and short-range nature of the charge correlations at high temperature.\cite{wang_high-temperature_2020}
These RIXS spectra display multiple excitation modes below $\sim$$100$~meV. 
Exemplary spectra at $\mathbf{q}=(H,0)$  with $H=0.15$ and $0.23$  are shown in Figures~\ref{fig:fig2}a--b.
Inelastic signals near 50 meV and 70 meV have been identified as bond-buckling and bond-stretching phonon modes, respectively (see Figure~\ref{fig:fig1}a), and their assignments have been established through INS,\cite{pintschovius_lattice_1991, mcqueeney_anomalous_1999, mcqueeney_mixed_2001, park_evidence_2014} IXS,\cite{fukuda_bond_2005, park_evidence_2014, huang_quantum_2021} first-principles calculations,\cite{giustino_small_2008} and photoemission experiments.\cite{zhou_multiple_2005}
Although multiple phonon modes are resolved by INS and IXS, RIXS primarily detects those with strong electron-phonon coupling.\cite{devereaux_directly_2016}
Excitations near 20 meV have also been observed in recent RIXS studies on La$_{1.875}$Sr$_{0.125}$CuO$_4$,\cite{martinelli_decoupling_2025} and YBa$_2$Cu$_3$O$_{7-\delta}$.\cite{arpaia_dynamical_2019} 
Thus, the entire low-energy line shape is fitted using five components: (1) elastic scattering (resolution-limited Voigt), (2) charge excitations (Gaussian), (3,4) two phonon modes (resolution-limited Voigt), and (5) background (second-order polynomial). These five components are shown in Figures~\ref{fig:fig2}a--b. More details of the fitting procedure are provided in Supplemental Note 2.
In the following text, we refer to the three inelastic modes as 20-meV, 50-meV, and 70-meV modes, respectively.

\begin{figure}[!t]
    \centering
    \includegraphics[width=0.99\linewidth]{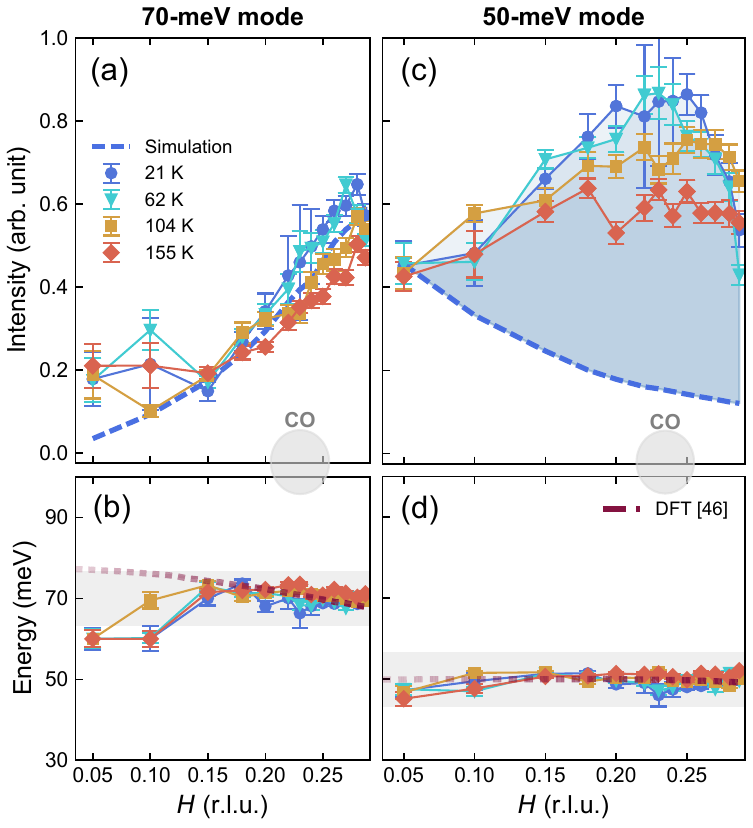}
    \caption{Momentum dependence of the 50- and 70-meV modes (a, b) Intensity and dispersion of the 70-meV mode. (c, d) Intensity and dispersion of the 50-meV mode. The blue shaded area indicates the additional excitation intensity relative to the phonon simulation. The dashed lines in (a, c) are obtained from the phonon simulation (see main text); the grey shaded areas in (b, d) are guides to the eye, and the dashed lines are scaled density functional theory (DFT) results from Ref.\cite{giustino_small_2008} CO, charge order.}
    \label{fig:fig3}
\end{figure}

The dispersion relations of the 70- and 50-meV modes are shown in Figures~\ref{fig:fig3}b and \ref{fig:fig3}d, respectively. Both modes display weak or moderate dispersion with no clear softening effect around the charge ordering wave vector, in qualitative agreement with previous INS and IXS studies.\cite{pintschovius_lattice_1991, mcqueeney_anomalous_1999, mcqueeney_mixed_2001, park_evidence_2014,fukuda_bond_2005, park_evidence_2014, huang_quantum_2021}
The peak amplitudes of the 70- and 50-meV modes are shown in Figures~\ref{fig:fig3}a and \ref{fig:fig3}c as a function of momentum transfer.

An enhancement of intensity is observed around the charge ordering wave vector for the 50-meV excitation, which decreases with increasing temperature. This behavior significantly deviates from the temperature evolution of a phonon mode.
To further understand this observation, we calculate the momentum-dependent RIXS intensity of the bond-stretching and bond-buckling phonon modes, respectively, and plot the results as dashed lines (details of the phonon simulation are described in the Supplemental Information). 
The calculation shows good agreement with the experimental data for the 70-meV mode, which corresponds to the energy scale of the bond-stretching phonon mode.
By contrast, the simulation substantially underestimates the excitation spectral weight around the charge ordering wave vector for the 50-meV excitation. 
Note that our calculated RIXS intensities for the bond-stretching and bond-buckling phonons are in qualitative agreement with previous results.\cite{devereaux_directly_2016} Quantitative differences may arise from the different electronic band structures associated with the different cuprate systems and doping levels considered in the two calculations.

\begin{figure*}[t]
    \centering
    \includegraphics[width = 0.95\linewidth]{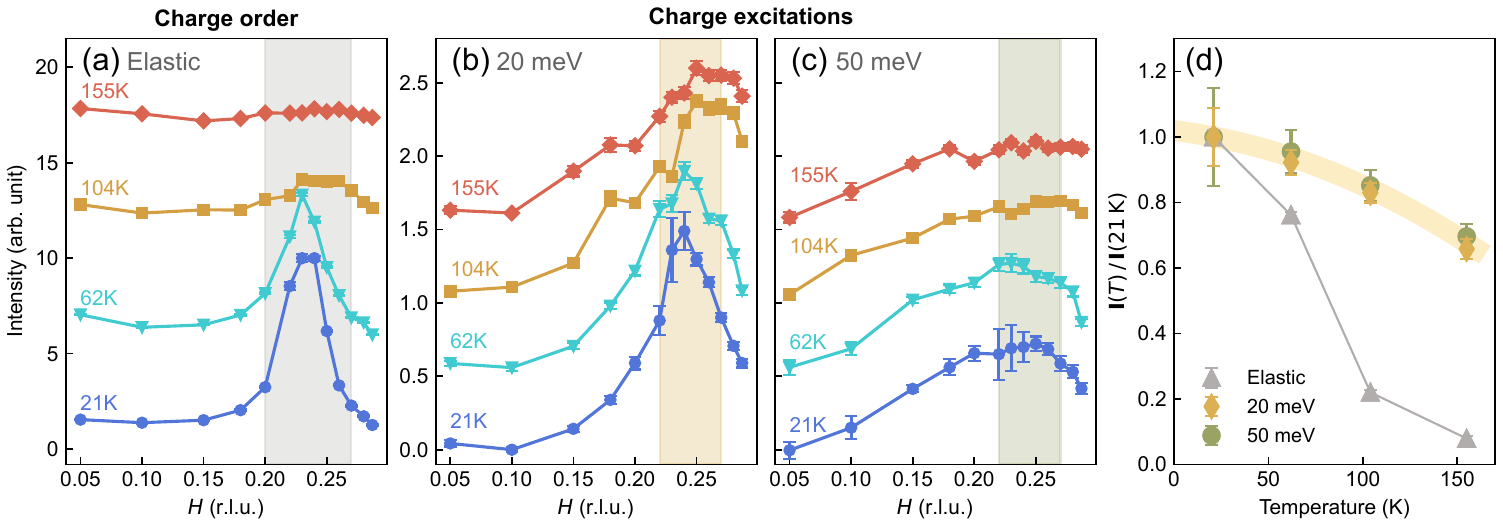}
    \caption{Temperature dependence of charge order and its fluctuations (a) Elastic scattering intensity (peak amplitude) along $(H,0)$ versus temperatures as indicated. (b) Inelastic intensity of the 20-meV mode. (c) The amplitude of the subtracted 50-meV mode for the same temperatures as in (a, b). All data in (a--c) have been given an arbitrary shift to enhance visibility. (d) Integrated intensities of elastic and inelastic signals in (a--c) as functions of temperature. The integration windows are indicated by the shaded areas. For comparison, the intensities are normalized to the value at the base temperature (21~K). The orange curve is a guide to the eye.}
    \label{fig:fig4}
\end{figure*}

In Figure~\ref{fig:fig4}, we summarize the temperature dependence. The elastic scattering intensity as a function of momentum transfer $(H,0)$ is shown in Figure~\ref{fig:fig4}a for the temperatures indicated. 
Similarly, in Figure~\ref{fig:fig4}b we plot the amplitude of the 20-meV excitation as a function of momentum transfer for the same temperatures. 
We stress that the elastic and the inelastic components display distinct temperature dependencies.
As mentioned before, the simulated bond-buckling phonon intensity does not fully account for the observed spectral weight of the 50-meV mode.
To understand this observation, we extract the additional excitation intensity by subtracting the simulation from the observed amplitude -- see Figure~\ref{fig:fig4}c.
In the following paragraphs, we refer to this as the \textit{subtracted} amplitude of the 50-meV mode.
Note that the labels `20-meV' and `50-meV' denote the dominant energy windows used to extract the charge-excitation spectral weight, and should not be interpreted as two independent flat charge-excitation branches.

The elastic and inelastic components are integrated around the peak center over a momentum window as indicated by the shaded area in Figure~\ref{fig:fig4}(a--c). 
The resulting integrated intensities as a function of temperature are shown in Figure~\ref{fig:fig4}d. The elastic signal decreases more rapidly with increasing temperature compared to the 20-meV excitation. 
We also stress that the subtracted 50-meV mode displays a temperature dependence similar to that of the 20-meV mode.
This could suggest a common origin for these two components.
Note that the intensity of the 70-meV excitation also exhibits a slight increase upon cooling at $H \gtrsim q_\text{CO}$, which may be further enhanced at a larger momentum transfer, as indicated by a previous Cu $L$-edge RIXS study.\cite{wang_charge_2021}

\begin{figure}[!t]
    \centering
    \includegraphics[width=\linewidth]{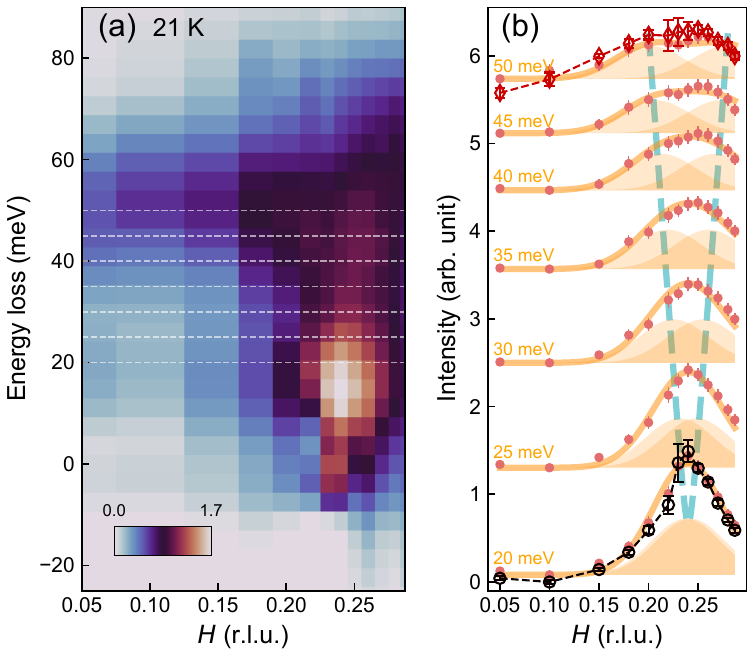}
    \caption{RIXS intensity and constant-energy cuts at 21~K (a) RIXS intensity map, with elastic and background signal subtracted, as a function of energy loss and momentum transfer. Horizontal white dashed lines indicate the energies of the cuts shown in (b). (b) Momentum dependence of RIXS intensity at the indicated energy losses. The energy integration window is 5~meV. 
    For comparison, intensities shown in Figure~\ref{fig:fig4} are overlaid: Black circles are amplitude of the 20-meV mode, while red diamonds are the phonon-signal-subtracted amplitude of the 50-meV mode. Orange curves are double-Gaussian fits, with the two components indicated by the shaded areas. Blue dashed lines trace the fitted peak positions as the energy increases.}
    \label{fig:fig5}
\end{figure}

An alternative approach to analyze the data is to subtract the elastic line profile and the background signal. In Figure~\ref{fig:fig5}a, we present the experimental RIXS intensity map at 21~K after subtracting these two contributions.
The corresponding momentum-dependent intensities at constant energy loss are shown in Figure~\ref{fig:fig5}b. For comparison, the amplitudes of the 20-meV mode (black circles) and the phonon-signal-subtracted amplitude of the 50-meV mode (red diamonds) are overlaid, demonstrating consistent results from the two analysis approaches.
The signal weakens and broadens in momentum as the energy increases. 
At higher energies ($\gtrsim40$~meV), a two-peak feature can be discerned.
Therefore, we applied a double-Gaussian fit model to capture the broadening and splitting, as illustrated by the orange shaded area in Figure~\ref{fig:fig5}b. 

\begin{figure}[!t]
    \centering
    \includegraphics[width=\linewidth]{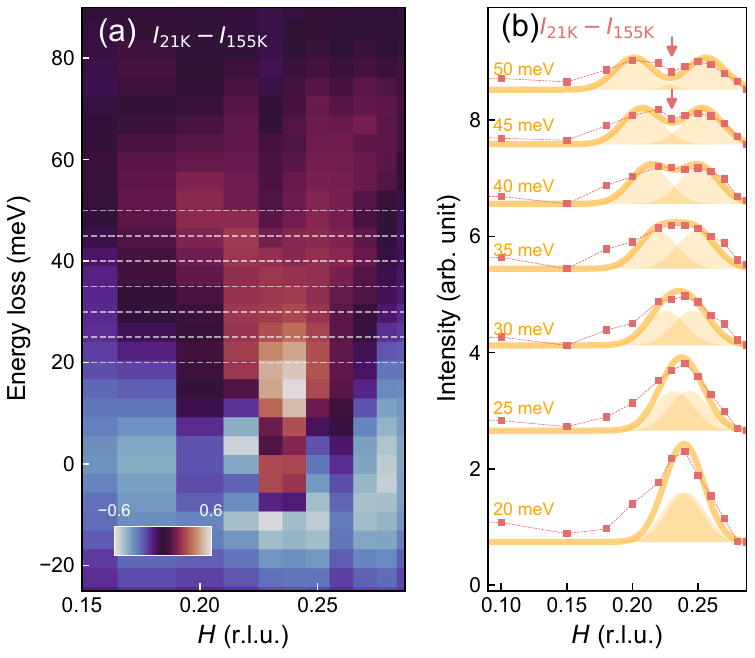}
    \caption{Temperature-difference analysis of the inelastic RIXS intensity (a) Difference between the inelastic RIXS intensities measured at 21~K and 155~K after subtraction of the elastic and background signals, $I^{\mathrm{inelastic}}_{21\,\mathrm{K}}-I^{\mathrm{inelastic}}_{155\,\mathrm{K}}$. Horizontal dashed lines indicate the energies of the cuts shown in (b). (b) Momentum dependence of the intensity difference at the indicated energy losses. Orange curves are double-Gaussian fits, and the shaded areas denote the two Gaussian components. The dip near $H\sim0.24$ above 40~meV reveals two dispersing branches.}
    \label{fig:fig6}
\end{figure}

We further justify the non-phonon origin of the dispersive feature without relying on the calculated phonon intensity. After removing the elastic signal and background, we subtract the inelastic intensity measured at 155~K from that measured at 21~K. The difference map, $I^{\mathrm{inelastic}}_{21\,\mathrm{K}}-I^{\mathrm{inelastic}}_{155\,\mathrm{K}}$, and its constant-energy cuts are shown in Figures~\ref{fig:fig6}a and \ref{fig:fig6}b, respectively. The 50- and 70-meV phonons are expected to change only marginally between these temperatures. The dispersive feature that remains in the difference therefore provides direct, model-independent evidence for a non-phonon excitation. Above 40~meV, the cuts develop a dip near $H\sim0.24$, consistent with two dispersing branches.\\

\TIPsection{DISCUSSION}
Our key observation is the existence of low-energy excitations that exhibit a temperature dependence distinct from what is expected for phonons, and also different from the charge order. 
As shown in Figure~\ref{fig:fig3}c, the intensity of the 50-meV mode decreases with temperature, contrary to phonons' behavior.
This intensity also decreases more slowly than that of the charge order, as illustrated in Figure~\ref{fig:fig4}d.
Therefore, we propose that a significant portion of the inelastic signals does \textit{not} originate from phonons. This claim is further supported by the discrepancy between the phonon simulation and the 50-meV mode.
A prevalent interpretation is that these low-energy excitations are the manifestation of charge fluctuations. In fact, phonon anomalies near the charge ordering wave vectors have been interpreted as evidence of charge fluctuations.\cite{li_multiorbital_2020, chaix_dispersive_2017, lee_spectroscopic_2020} Furthermore, their quantum critical nature in cuprates has previously been discussed.\cite{huang_quantum_2021, arpaia_signature_2023}
In this work, we focused on the excitations well within the ordered state.
An outstanding question is the propagating nature of these charge fluctuations. 
Extracting the dispersion of charge fluctuations has proven challenging due to the overlap with elastic signal and phonon contributions. A direct observation of dispersive charge excitations has thus far not been realized, hindering the understanding of the microscopic charge ordering mechanism.

The high energy resolution of O $K$-edge RIXS enabled us to filter out the elastic scattering from, for example, charge fluctuations.
By modeling the phonon modes with strong electron-phonon coupling, it is also possible to separate the spectral weight contributions from phonons and charge fluctuations. 
This step in our analysis relies on the calculation of the momentum-dependent cross section of the phonon modes, a method that has not been employed in this context before.
Although higher momentum transfer is not accessible due to the kinematic constraints of x-rays at O $K$-edge, the enhanced energy resolution compared to Cu $L$-edge RIXS is crucial for the extraction of charge excitations.

\begin{figure}[!t]
    \centering
    \includegraphics[width=0.95\linewidth]{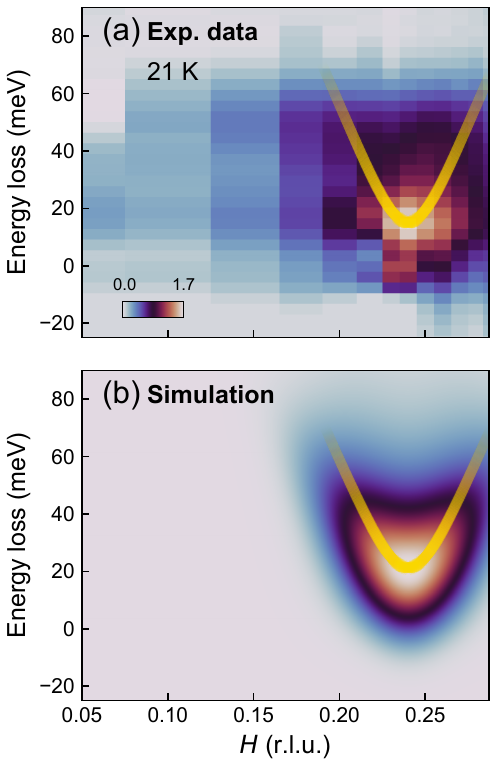}
    \caption{Charge-excitation spectral weight and simulation (a) Experimental RIXS intensity after subtraction of the elastic signal, background, and simulated phonon intensity. The yellow curve is the modeled charge-excitation dispersion at 21 K. (b) Simulated charge dynamical structure factor $S(\mathbf{q},\omega)$ using the dispersion parameters obtained from the fits in Figure~\ref{fig:fig5}b.}
    \label{fig:fig7}
\end{figure}

By eliminating the elastic scattering and the phonon contributions from the RIXS spectra, we are able to isolate the signal stemming from charge fluctuations. 
If we presume it to be a long-range dispersive collective mode, we can assume the following dispersion relation:
$\omega_{\mathrm{cf}} = \sqrt{\Delta^2 + v_c^2|\mathbf{q}-\mathbf{q}_{\mathrm{CO}}|^2}$.
The energy gap $\Delta = 18$~meV is estimated from the spectral fit shown in Figure \ref{fig:fig2}b. The excitation velocity (stiffness) $v_c$ around the charge ordering wave vector $\mathbf{q}_\text{CO}\approx (0.23,0)$ is determined by a global double-Gaussian fit to the momentum-dependent cuts of the 21-K data in Figure~\ref{fig:fig5}b, with the two peak centers constrained by this dispersion relation. The fitted dispersion is overlaid on the phonon-subtracted charge-excitation intensity in Figure~\ref{fig:fig7}a.

To model the charge excitation, we use an ansatz that captures its bosonic character:
\begin{equation}
	\chi(\mathbf{q}, \omega) = \frac{1}{\omega_\mathrm{cf}^2 - (\omega + \mathrm{i}\Gamma)^2},
\end{equation}
where $\chi$ denotes the charge susceptibility, and $\Gamma$ is the dissipation parameter. The dynamical charge structure factor $S(\mathbf{q},\omega)$ is given by
\begin{equation}
	S(\mathbf{q},\omega) = \left(1 - e^{-\frac{\hbar \omega}{k_\mathrm{B}T}}\right)^{-1}\mathrm{Im}\chi(\mathbf{q},\omega),
\end{equation}
where $k_\mathrm{B}$ is the Boltzmann constant and $T$ is the temperature. As shown in Figure~\ref{fig:fig7}, the calculated structure factor is in qualitative agreement with the experimental result.

Our data suggest a charge fluctuation velocity (stiffness) of $v_c=(800\pm 200)$~meV~\AA~($\sim 1.3\times 10^{5} \,\mathrm{m/s}$), which is comparable to that of the paramagnons ($v_s\sim800$~meV~\AA).\cite{coldea_spin_2001}
In fact, a RIXS study on Hg-based cuprates suggested that charge excitations may extend to around 200~meV, reaching the bandwidth of the paramagnon.\cite{yu_unusual_2020}
The comparable energy scales likely originate from the intimate coupling between the spin and charge orders.

Previous RIXS studies have estimated the charge excitation velocity based on indirect observations, e.g., anomalies in the phonon intensity away from $\mathbf{q}_\text{CO}$,\cite{lee_spectroscopic_2020, chaix_dispersive_2017, li_multiorbital_2020} or by modeling the charge excitation intensity below the bond-buckling phonon energy.\cite{huang_quantum_2021} 
The velocity values derived from these estimations exhibit large discrepancies, ranging from 450 to 1300~meV~\AA. 
This large uncertainty has hindered our fundamental understanding of the related quasiparticle and its impact on the normal state and superconducting properties.
Here, by carefully modeling the RIXS phonon responses, we isolate the charge excitation contributions and directly establish a dispersive charge excitation mode extending to at least 60 meV (Figure~\ref{fig:fig7}a). 
Similar attempts but with different strategies have been made, for example, by subtracting the scattering intensity along the diagonal $(H,H)$ direction,\cite{arpaia_signature_2023} or by assuming the momentum-dependent phonon contribution as a simple polynomial background.\cite{yu_unusual_2020} Although these works revealed the existence of charge fluctuations, the dispersion has never been determined.

In contrast to cuprate systems with weaker charge order or closer to the quantum critical point (QCP), where charge excitations appear as a continuum,\cite{lee_spectroscopic_2020,huang_quantum_2021,arpaia_signature_2023} Eu-LSCO with $x=0.125$ exhibits well-developed charge correlations (see Supplemental Note 1). This makes it possible to extract the characteristic energy-momentum dependence of the charge-excitation spectra. The observed dispersion therefore provides an estimate of the interaction scale governing charge ordering. Nevertheless, the charge excitations in Eu-LSCO are still damped, likely because of the presence of low-energy particle-hole excitations that provide a decay channel.

The quantitative determination of charge excitation dispersion is crucial for establishing the underlying interactions that involve both the spin and charge degrees of freedom. The comparable energy scales of collective charge and spin excitations suggest a strong coupling realized in the form of fluctuating stripes.\cite{kivelson_how_2003}
As pairing does not coexist with local antiferromagnetic order, charge fluctuations may enhance Josephson coupling between neighbouring stripes and promote superconductivity.\cite{kivelson_electronic_1998} 
In this context, collective charge excitations may contribute constructively to superconductivity.
In fact, numerical calculations have shown that electron-electron (exchange) interactions that mediate pairing could also be responsible for charge ordering.\cite{davis_concepts_2013,zheng_stripe_2017,huang_numerical_2017}
This theoretical insight aligns with our observation of a common energy scale for charge and spin excitations, as well as the superconducting gap. Besides superconductivity, charge fluctuations may also profoundly influence the normal state properties. It has recently been suggested that electron dissipation due to scattering from charge fluctuations leads to the strange metal behavior.\cite{arpaia_signature_2023} 
Such a scenario can be further verified by establishing the evolution of the charge excitation
dispersion and normal state electronic properties throughout the phase diagram.

\TIPsection{CONCLUSION}

By integrating phonon cross-section calculation and high-resolution RIXS experiments, our study successfully isolated weak charge excitations and established their dispersion and energy scale in cuprates. This approach can be extended to other strongly correlated materials, including nickelate and iron-based superconductors, where charge-spin coupling is significant.\cite{li_stripes_2017,chen_charge_2023,bu_observation_2021,Sun_signatures_2023,zhang_intertwined_2020} As such, our study lays the groundwork for uncovering universal behaviors that link charge dynamics to emergent phases in complex quantum materials.

\TIPsection{REFERENCES}

\renewcommand{\bibnumfmt}[1]{#1.}
\renewcommand{\doi}[1]{DOI:\href{https://doi.org/#1}{#1}}

\TIPsection{FUNDING AND ACKNOWLEDGMENTS}

We thank Zheng-Cheng Gu, Di-Jing Huang, Hsiao-Yu Huang, Yao Wang, and David Leboeuf for insightful discussions. X.H., K.v.A., and J.~Chang thank the Swiss National Science Foundation under Project No.~200021\_188564. Y.Y. and Q.W. are supported by the Research Grants Council of Hong Kong (CUHK 24306223, CUHK 14305926), the Guangdong Provincial Quantum Science Strategic Initiative (GDZX2401012), CUHK Direct Grant (4053671, 4053791), and the 1+1+1 CUHK-CUHK(SZ)-GDSTC Joint Collaboration Fund (2025A0505000079). I.B. and L.M. acknowledge support from the Swiss Government Excellence Scholarship under project numbers ESKAS-Nr: 2022.0001 and ESKAS-Nr: 2023.0052. J.~Choi acknowledges financial support from the National Research Foundation of Korea (NRF) funded by the Korean government (MSIT) through Sejong Science Fellowship (Grant No.~RS-2023-00252768). Y.S. is supported by the Wallenberg Academy Fellows through grant 2021-0150. M.-H.~J.'s work at LNCMI was supported by the French Agence Nationale de la Recherche (ANR) under reference ANR-25-CE30-2817. We acknowledge Diamond Light Source for providing beam time on beamline I21 under proposal MM29011. We also acknowledge ESRF for beam time allocation on beamline ID32 under proposal HC-6093.

\TIPsection{AUTHOR CONTRIBUTIONS}

J.~Chang and Q.W. conceived the project. S.P., T.T., and H.T. grew the Eu-LSCO $x=0.125$ single crystals. D.Z., Y.Z., T.W., and X.C. grew the Eu-LSCO $x=0.26$ single crystals. K.v.A., J.~Choi, Y.S., M.G.F., K.J.Z., J.~Chang, and Q.W. performed RIXS experiments at I21. X.H., C.F., L.C., M.F., I.V., F.Y., and K.V. performed RIXS experiments at ID32. Data analysis and phonon simulations were carried out by X.H. with assistance from Y.Y., L.M., and I.B. M.-H.~J. helped with interpretation of the results. The manuscript was written by X.H., J.~Chang, and Q.W. with input from all authors.

\TIPsection{DATA AND CODE AVAILABILITY}

Data are available from the corresponding authors upon reasonable request.

\end{document}